\documentclass{aastex}

\newcommand{\kms}{\mbox{km s$^{-1}$}}

\newcommand{\Msun}{\mbox{\,M$_{\odot}$}}
\newcommand{\Lsun}{\mbox{L$_{\odot}~$}}
\newcommand{\um}{\mbox{$\mu$m }} 
\newcommand{\co}[2]{\mbox{CO($#1$$\rightarrow$$#2$)}}

\newcommand{\cdo}[2]{\mbox{C$^{18}$O($#1$$\rightarrow$$#2$)}}
\newcommand{\hcop}[2]{\mbox{HCO$^+$($#1$$\rightarrow$$#2$)}}
\newcommand{\sio}[2]{\mbox{SiO($#1$$\rightarrow$$#2$)}}
\newcommand{\skipthis}[1]{}

\newcommand{\degs}{$\,^\circ$}

\shortauthors{Guzm\'an et al.}
\shorttitle{Molecular outflows towards IRAS 16562$-$3959 jet}
\usepackage{amssymb,amsmath,natbib,graphicx,rotating,wasysym}
\begin{document}


\title{A hot molecular outflow driven by the ionized jet 
associated with IRAS 16562$-$3959}

\author{Andr\'es E. Guzm\'an\altaffilmark{1}, Guido Garay\altaffilmark{1},
\and Kate J. Brooks\altaffilmark{2} \and Jill Rathborne\altaffilmark{1,2} \and 
Rolf G\"usten\altaffilmark{3}}


\altaffiltext{1}{Departamento de Astronom\'{\i}a, Universidad de Chile,
  Camino el Observatorio 1515, Las Condes, Santiago,
  Chile}\altaffiltext{2}{Australia Telescope National Facility, CSIRO
  Astronomy and Space Science, P.O. Box 76, Epping 1710 NSW,
  Australia}\altaffiltext{3}{Max-Planck-Institut f\"ur Radioastronomie, Auf
  dem H\"ugel 69, 53121 Bonn, Germany}

\begin{abstract}

We report molecular line observations in the CO J=3$\rightarrow$2,
6$\rightarrow$5 and 7$\rightarrow$6 transitions, made using the Atacama
Pathfinder Experiment Telescope (APEX), toward the massive and dense core
IRAS 16562$-$3959.  This core harbors a string of radio sources thought to
be powered by a central collimated jet of ionized gas.  The molecular
observations show the presence of high velocity gas exhibiting a
quadrupolar morphology, most likely produced by the presence of two
collimated outflows.  The southeast-northwest molecular outflow is aligned
with the string of radio continuum sources, suggesting it is driven by the
jet. We find that the excitation temperature of the gas in the SE-NW
outflow is high, with values of 145 and 120 K for the blueshifted and
redshifted lobes, respectively.  This outflow has a total mass of 1.92
\Msun, a total momentum of $\sim 89$ \Msun~\kms~ and an averaged momentum
rate of $\sim 3.0\times10^{-2}$ \Msun~\kms~ yr$^{-1}$, values
characteristics of flows driven by young massive stellar objects with high
luminosities ($L_{\rm bol} \sim 2\times10^4$ \Lsun).  Complementary data
taken with the Atacama Submillimeter Telescope Experiment (ASTE) in high
density and shock tracers support the picture that IRAS 16562$-$3959 is an
accreting young massive star associated with an ionized jet, which is the
energy source of a molecular outflow.

\end{abstract}  


\vfill\eject

\section{INTRODUCTION}

IRAS 16562$-$3959 (also G345.4938+01.4677) is a mid-infrared (MIR) source,
with \emph{MSX} \textit{(Midcourse Space Experiment)} colors characteristics of
massive young stellar objects (MYSOs) \citep{Lumsden2002MNRAS}.
Observations of dust continuum emission at 1.2 mm, made with SIMBA at the
SEST telescope, show that IRAS 16562$-$3959  is
associated with a massive ($1.3\times10^3 M_{\odot}$) and dense
($9\times10^5$ cm$^{-3}$) molecular core \citep{Lopez2011AA}. Maser
emission has been detected in transitions of OH
\citep{Caswell1998MNRAS297215,Caswell2004MNRAS}, giving further support to
the hypothesis that IRAS 16562$-$3959 is a young high-mass star forming region.

Recently, \citet{Guzman2010ApJ} reported the detection toward IRAS
16562$-$3959 of a string of radio continuum emission consisting of a bright
compact central object and four outer lobes.  They argue that the continuum
emission from the central object corresponds to free-free emission from a
thermal jet whereas the radio emission from the lobes correspond to thermal
emission arising in shocks generated from the interaction of the collimated
wind with the surrounding medium.  Assuming that IRAS 16562$-$3959 is
located at a distance of $1.6\pm0.1$ kpc \citep[$V_\textrm{LSR}=-12.6$
  km/s,][]{Urquhart2008AA}\footnote{The near/far kinematic distance
  ambiguity was resolved in \citet{Faundez2004AA}}, the total far-infrared
luminosity, computed using the observed IRAS fluxes
\citep[see][]{Casoli1986AA}, is $\sim7.0\times10^4$ \Lsun.  This implies
that IRAS 16562$-$3959 harbors the most luminous MYSO known to date
associated with an ionized jet.

Mid-infrared images taken from the {\sl Spitzer Space Telescope}-GLIMPSE
survey data \citep{Benjamin2003PASP} show bright emission associated with
the jet in all IRAC bands. In particular, at 8.0 \um there is extended
emission roughly aligned with the jet axis. Further, there appears to be an
excess of emission at 4.5\um, which is thought to be indicative of shock
activity \citep{Cyganowski2008AJ,Chambers2009ApJS}.  The GLIMPSE data also
show that there is at least one more young stellar object (YSO) within the
core, located about 15\arcsec\ northeast of the jet, that is not detected
at 2 \um and hence likely to be deeply embedded.

In this paper we present molecular line observations toward IRAS 16562$-$3959 
in the CO($3\rightarrow2$), CO($6\rightarrow5$), and
CO($7\rightarrow6$) transitions, made using APEX, which have revealed the
presence of two high velocity bipolar molecular outflows. Their observed
and derived characteristics suggest that the driving sources are young,
luminous protostellar objects. One of
the outflows is aligned in the same direction as the string of radio
sources strongly suggesting that it is being driven by the thermal radio jet.

\section{OBSERVATIONS}

The observations were made using the 12-m Atacama Pathfinder Experiment
Telescope (APEX) located at Llano de Chajnantor, Chile. A detailed
description of APEX and its performance are given by \citet{Gusten2006AA}.
The observed transitions and basic observational parameters are summarized
in Table~\ref{tbl-obs}. Columns 1 and 2 give, respectively, the observed
transition and line frequency. Columns 3, 4 and 5 give the telescope used,
the half-power beam width and main beam efficiency at the observed
frequency. Columns 6 to 9 give, respectively, the number of positions
observed, the angular spacing, the channel width, and the resulting rms
noise in antenna temperature, for each of the observed transitions.

The CO($3\rightarrow2$) observations were made during June, 2008 and July,
2009.  The front-end consisted of a single pixel heterodyne SiS receiver
operating in the 275-370 GHz band \citep{Risacher2006AA}. For the back-end
we used the APEX Fast Fourier Transform Spectrometer \citep{Klein2006AA}
with a bandwidth of 1 GHz and 2048 frequency channels. The velocity
coverage was $\sim 870$ \kms\ centered at $-13.6$ \kms\ V$_\textrm{LSR}$ with a channel
width of 0.42 \kms. In this transition the emission was mapped within a
region of $100\arcsec\times100\arcsec$ in size, with 20\arcsec\ angular
spacing (roughly full beam spacing), centered on the position of the radio
source. System temperatures were typically 270 K. The pointing accuracy is
$2-3$\arcsec\ calibrated towards the source RT-Sco
($\alpha=17^h$03$^m$32.6$^s$, $\delta=-36$\degs55\arcmin14\arcsec) and the
uncertainty in the absolute flux scale is $\sim 10\%$.  The observations
were performed in position-switching mode under good atmospheric conditions
($\tau\sim0.07$) using $\alpha=16^h$58$^m$15$^s$ and
$\delta=-39$\degs00\arcmin00\arcsec as off-position.  The integration time
on-source in each position was $\sim30$ seconds, resulting in an rms noise
of typically 0.1 K in antenna temperature per channel.

The CO($6\rightarrow5$) and CO($7\rightarrow6$) observations were made
during August, 2009, under good weather conditions (0.35-0.45 mm of water
vapor, corresponding to $0.5<\tau<0.8$) using an off-position located at
$\Delta\alpha=+6^m$. The pointing calibration source was NGC6334, attaining
an accuracy of $\sim2$\arcsec. The front-end consisted of a dual heterodyne
SIS receiver array of 7 pixels operating in the 600-720 GHz and 750-950 GHz
atmospheric windows, known as the CHAMP$^+$ receiver
\citep{Gusten2008SPIE}.  In these two transitions we mapped the emission
within a region of $110\arcsec\times90\arcsec$ in size, with
4\arcsec\ angular spacing. The back-ends were operated with 2.4 GHz total
bandwidth with 2048 channels each.

We complement the above data with additional spectra taken toward the radio
source position in the \sio{8}{7}, \hcop{4}{3} and \cdo{3}{2} lines. The
\sio{8}{7} observations were performed using APEX with the same APEX FFT
spectrometer.  The integration time was $\sim450$ seconds giving an rms
noise of 0.04 K.

The \hcop{4}{3} and \cdo{3}{2} spectra were taken as part of a larger
mapping of the core using the ASTE 10 m submillimeter telescope
\citep{Ezawa2004SPIE} during June-July, 2010, under good atmospheric conditions
($\tau\sim0.08$). The front-end consisted in a heterodyne single pixel
SiS receiver operating between 324 and 372 GHz. As back-end we used the MAC
XF-type digital spectro-correlator with 1024 channels. For the \cdo{3}{2}
line we used a total bandwidth of 512 MHz, giving a velocity resolution of
about 0.43 \kms, whereas for the \hcop{4}{3} transition we used a 125 MHz
bandwidth, giving a velocity resolution of $\sim 0.11\,$\kms.  A detailed
description of these observations will be given elsewhere.

{\section{RESULTS}\label{section-results}} 

\begin{figure}
\includegraphics[angle=-90,width= \textwidth ]{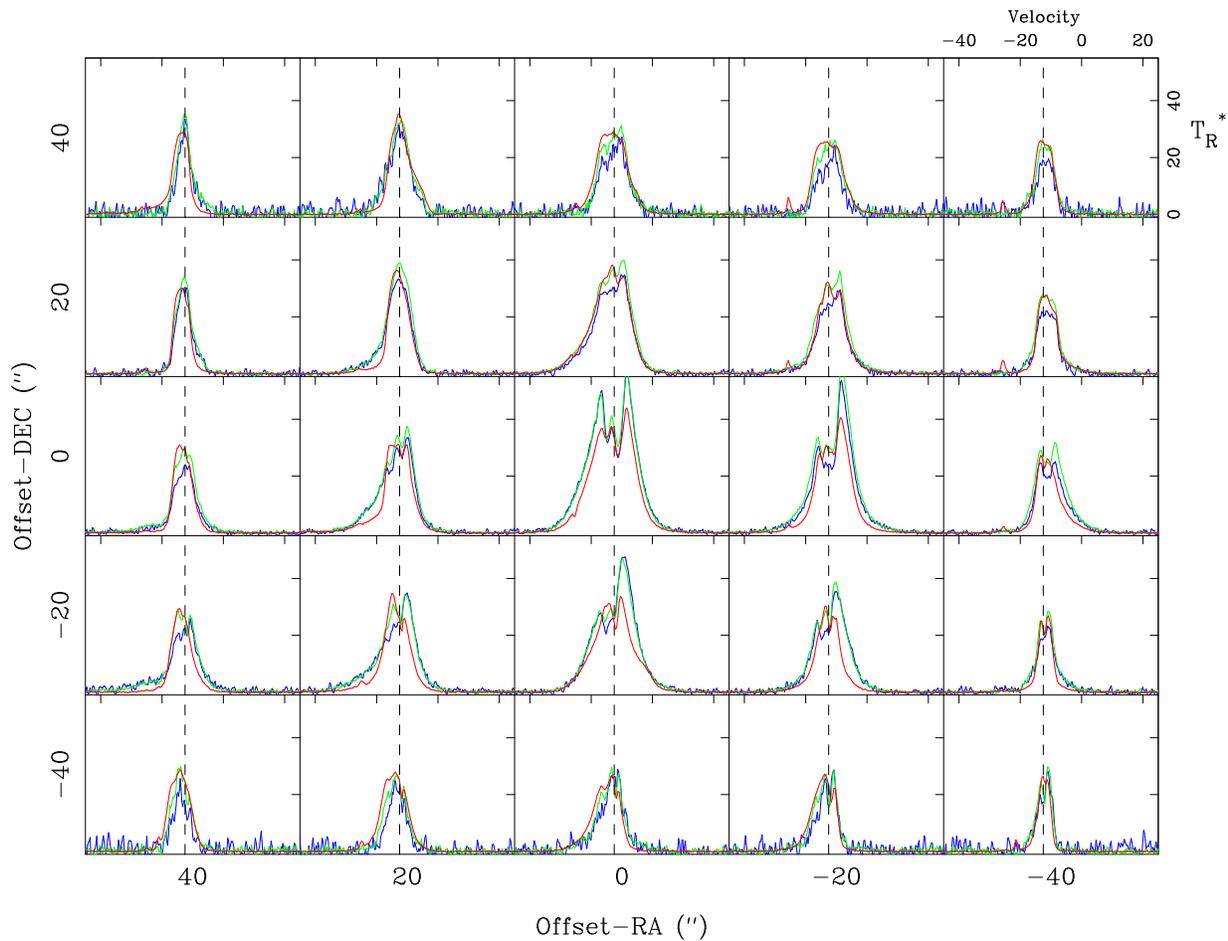}
\caption{
 Spectral grid of the \co{3}{2} (red), \co{6}{5}
  (green), and \co{7}{6} (blue) emission observed towards IRAS 16562$-$3959.
  The grid spacing is 20\arcsec.  Offsets are from the radio source
  reference position at $\alpha_{2000} = 16^{\rm h}59^{\rm m}41.61\fs0,
  \delta_{2000} = -40\arcdeg 03\arcmin 43\arcsec$.  Velocity scale ranges
  from $-45$ to $25$ \kms. The dashed line in each spectrum marks the
  ambient velocity adopted of $-12.5$ \kms.
\label{fig-326576}}
\end{figure}

Figure \ref{fig-326576} shows a grid of the CO spectra in the three
observed transitions within a region of 100\arcsec $\times$ 100\arcsec\ in
size, with 20\arcsec\,spacing. The spectra of the \co{6}{5} and \co{7}{6}
transitions were convolved to match the angular resolution of the \co{3}{2}
spectra.  Clearly seen in this figure is the presence of broad and strong
wing emission across the molecular core.

Figure~\ref{fig-hcomas} shows the \hcop{4}{3}, \cdo{3}{2} and \sio{8}{7}
spectra observed toward the peak position of IRAS 16562$-$3959. The \hcop{4}{3}
spectrum shows a double-peaked line profile, with a bright blue-shifted peak
at $-14.0\pm0.2$ \kms\ and a weaker red-shifted peak at $-10.1\pm 0.2$
\kms. On the other hand, the \cdo{3}{2} line profile shows a symmetric
single component with a peak line center velocity of $-12.5$ \kms. We also
note that the \hcop{4}{3} and the \sio{8}{7} spectra show prominent high
velocity wings.

The $^{12}$CO profiles exhibit redshifted wing emission mainly towards the
west and south, up to an LSR velocity of $14$ \kms, whereas blueshifted
wing emission is seen mainly towards the east and north, up to an LSR
velocity of $-46$ \kms.  The full velocity range of the wing emission
observed is then $\sim 60$ \kms. The radial flow velocity is defined as
$\mid {\rm v}_{\small \textrm{lsr}}-{\rm v}_{\textrm{amb}} \mid $, where
${\rm v}_{\rm amb}$ is the systemic velocity of the ambient gas.  For the
later we adopt a value of $-12.5$ \kms, corresponding to the peak velocity
of the \cdo{3}{2} line profile observed toward the central position of the
core. The maximum outflow velocity is then $\sim 33.5$ \kms\ towards the
blue and $\sim 26.5$ \kms\ towards the red.  Figure~\ref{fig-co-outflows}
shows contour maps of the wing emission in the three CO lines integrated
over the velocity range $-46<{\rm v}_{lsr}<-22.0$ \kms\ (blue contours) and
$-3<{\rm v}_{lsr}< 14$ \kms\ (red contours). The star marks the position of
the radio jet \citep{Guzman2010ApJ}. These velocity limits define the
ranges in which emission from the high-velocity outflow gas is seen. We
note that the blueshifted flow emission is slightly stronger and reaches
higher radial flow velocities compared to the redshifted emission, hence
the wider blueshifted velocity range.  The inner limits of the ranges were
selected as to leave out the contribution of the ambient cloud and chosen
symmetrically respect to the ambient cloud velocity.

The spatial distribution of the blueshifted and
redshifted emissions shows a quadrupolar morphology, with blueshifted
emission seen mainly toward the east and north and redshifted emission seen
toward the west and south. We propose that the quadrupolar morphology is
produced by the superposition of two bipolar outflows, one along a
southeast-northwest direction (hereafter the SE-NW flow) with a position
angle of $\sim 107^\circ$, and a second outflow aligned roughly in a
north-south (N-S) direction.  The collimation factors --- length divided by
width --- of the molecular outflows are not high: for the blueshifted and
redshifted lobes of the SE-NW outflow are $\gtrsim 2$ and $\sim 1$,
respectively, whereas for the lobes of the N-S outflow we found values
close to 1.  These values are similar to those derived for massive outflows
\citep{Beuther2002AA}.  We note, however, that the highest velocity gas
emission of the SE-NW outflow appears to be associated with more collimated
structures.

Figure \ref{fig-outjet} presents a contour map of the outflow emission in
the \co{6}{5} line overlayed with the 8.6 GHz radio continuum emission
observed toward IRAS 16562$-$3959 \citep{Guzman2010ApJ}. It appears that
the SE-NW outflow is associated with the string of radio sources, the peak
position of the blue-shifted and red-shifted lobes being symmetrically
displaced from the bright central radio source. The SE blueshifted and NW
redshifted lobes extend up to $\sim 27\arcsec$ and $\sim 32\arcsec$ from
the central radio source, respectively. The symmetry axis of the SE-NW
outflow is along a direction with a position angle of $\sim107$\degs,
roughly the same as the P.A. of the symmetry axis of the jet of $110$\degs.
As noted in \citet{Guzman2010ApJ}, the radio lobes and the jet are not
completely aligned, showing a small bending, which is also seen in the
SE-NW outflow.  Possible bending mechanisms of protostellar jets are  
discussed in \citet{Fendt1998AA}, but with the available data we can 
not discern between the various alternatives. 

The spectroscopic signatures of the \hcop{4}{3} and \cdo{3}{2} transitions
suggest that the bulk of the molecular gas toward IRAS 16562$-$3959 is
undergoing large-scale inward motions \citep[e.g.][]{Sanhueza2010ApJ}.
Infalling motions traced by optically thick molecular lines are expected to
produce line profiles showing blue asymmetry, whereas optically thin lines
are expected to exhibit symmetrical profiles \citep{Mardones1997ApJ}.

\section{ANALYSIS AND DISCUSSION}
\subsection{K$_s$-band evidence of shocked H$_2$}

Figure \ref{fig-2mass} presents an image of the Two Micron All Sky Survey
(2MASS, \citealt{Skrustskie2006AJ}) $K_s$-band emission across an
$8\arcmin\times8\arcmin$ region of the sky, centered near IRAS 16562$-$3959.
Clearly seen toward the center is diffuse emission along the SE-NW
direction extending by more than $1.5$\arcmin\ on each side of the bright
radio source.  The position angle of this diffuse $K_s$-band emission is
$\sim104^\circ$, a value similar to that of the SE-NW molecular outflow.
We suggest that the outer parts of this diffuse emission, which are at
greater distances from the central jet than the molecular lobes, are
tracing shocked H$_2$-2.12\,\um\ gas produced by an older episode of mass
ejection from the jet.

We note that the NIR (near infrared) emission detected close to the central
jet comes mainly from a region associated with the SE blueshifted lobe.
Much less emission is seen associated with the NW redshifted lobe (see also
Fig. 4 from Guzm\'an et al. 2010). A similar situation has been observed in
other bipolar outflows, for example in BHR 71 \citep{Bourke2002osp}.  Most
likely, the strong NIR emission from the eastern, closer to the jet,
feature correspond to scattered light from the inner walls of the outflow
cavity.
 
The second outflow detected toward IRAS 16562$-$3959 is roughly aligned in the
north-south direction, having a symmetry axis with a position angle of
$7^\circ$.  The blueshifted lobe extends up to $\sim21\arcsec$ to the north
of the jet and the redshifted lobe up to $\sim 17\arcsec$ to the
south. Interestingly, Fig.  \ref{fig-2mass} also shows the presence of
diffuse $K_s$-band emission roughly along the N-S direction but located
about 3.4\arcmin\ north and 2\arcmin\ south from the NS molecular outflow.
This emission exhibits a bow shock morphology, characteristic of HH
objects, most likely produced by the interaction of a collimated flow with
the ambient medium.  However, as noted in \citet{Guzman2010ApJ}, this
diffuse emission is bright in all mid-infrared \emph{MSX} bands. In
consequence, it is likely that this $K_s$-band emission has an important
NIR continuum component, in addition to the H$_2$-2.12\,\um\ line.

We suggest that there is an independent source of energy driving the NS
outflow, although from the available data we are not able to pinpoint its
location. \citet{Guzman2010ApJ} have already argued, on the basis that a
re-orientation of the central jet in $\sim90$\degs\ seems physically
unfeasible, for the presence of two high-mass YSOs driving the outflows.

\subsection{Parameters of the outflows}

To compute physical parameters of the molecular outflows we followed the
standard formalism described in \citet{Bourke1997ApJ},
\citet{Garden1991ApJ} and \citet{Goldsmith1999ApJ}, assuming that the high
velocity gas is optically thin and its excitational state can be described
by a single excitation temperature. If the transitions are sub-thermally
excited, then the derived excitation temperature would correspond to a
lower limit of the kinetic temperature of the outflowing gas.

A general discussion of the sources of errors have been given by
\citet{Margulis1985ApJ} and \citet{Cabrit1990ApJ}. The main sources of
error arise from the difficulty in determining the contribution to the
outflow in the velocity range of the ambient cloud, and not knowing the
flow inclination. To be conservative we adopt as velocity boundary between
the blue and red wing emission and the ambient emission the values of
$-22.0$ and $-3.0$ \kms, respectively.

{\subsubsection{Column densities and excitation temperatures}\label{sec-NandTex}}

The column density of CO molecules in the velocity  range $[v1,v2]$ is given by 
\begin{equation}
N_{{\rm CO}}=2.31\times10^{14}~  \frac{(T_{\rm ex}+h B/3k) }{1-\exp\left( -h \nu/kT_{\rm ex}\right)}\frac{\exp\left(E_J/k 
T_{\rm ex}\right)}{(J+1)}\int_{v1}^{v2} \tau_v dv\quad\text{cm}^{-2}~~,
\label{eq-columnCO}
\end{equation}
where $T_{\rm ex}$ is the rotational excitation temperature in K, $J$ is the
rotational quantum number of the lower state, $E_J=h B J (J+1)$ is the
energy of level $J$, where $B$ is the rotational constant of the CO
molecule.  The frequency of the \co{J$+1$}{J} transition is $\nu\approx2 B
(J+1)$, and $\tau_v$ is the opacity of the material moving at velocity $v$
(the latter measured in \kms).  Assuming a beam filling factor of 1, the
observed main-beam brightness temperature is related to the opacity by
\begin{equation}\label{eq-tauvsJa}
T^*_{\rm mb}(v)=[J_r(\nu,T_{\rm ex})-J_r(\nu,T_{\rm bg})](1-e^{-\tau_v})~~,
\end{equation}
where $J_r(\nu,T)$ is defined as
\[ J_r(\nu,T)=\frac{h \nu/k}{\exp(h\nu/kT)-1}~~. \]
In the optically thin limit,
\begin{equation}\label{eq-tauvsJ}
T^*_{\rm mb}(v)= [J_r(\nu,T_{\rm ex})-J_r(\nu,T_{\rm bg})]\tau_v~~.
\end{equation}

From the observations of the emission in the three transitions of CO it is
possible to determine both the excitation temperature and column density of
the outflowing gas using Eqs. \eqref{eq-columnCO} and \eqref{eq-tauvsJ},
and taking $T_{\rm bg}=2.7$\ K.  Figure \ref{fig-vels} presents plots of the
velocity integrated emission versus the $J$ quantum number from the
blueshifted and redshifted lobes of both outflows and from the ambient gas.
The velocity intervals of integration for the blueshifted and redshifted
emission are $[-46,-22]$ and $[-3,14]$\, \kms, respectively. The areas
over which the emission has been spatially integrated for the different
outflow lobes are shown in Fig. \ref{fig-outjet}.  In the positions in
which lobes overlap we assumed equal contributions from each lobe, except
at offset position ($\Delta \alpha=+20$\arcsec\, $\Delta
\delta=+20$\arcsec) where we assumed that the contribution to the
blueshifted emission from the NS and SE-NW outflows are in a 2:1 ratio.
Dashed lines show the results of the best fit to the observed integrated
emission assuming optically thin conditions and filling factors of 1.  The
derived CO column densities and excitation temperatures are given in Table
\ref{param-vels}. The parameters for the ambient cloud were derived using
the spectra observed at offset position ($\Delta\alpha=+40\arcsec$,
$\Delta\delta=-40\arcsec$), which does not show evidence of self-absorption
nor high velocity gas.

One of the less well known physical parameter of the gas in molecular
outflows is its kinetic temperature. Our observations of three lines of CO
allowed us to determine the excitation temperature of both outflows,
concluding that their kinetic temperatures are high.  The derived
excitation temperatures of the high-velocity gas are 145 and 120 K for the
blueshifted and redshifted lobes of the SE-NW outflow, and  76 and 88 K
for the blueshifted and redshifted lobes of the NS outflow. All these
temperatures are considerably higher than the ambient cloud temperature,
indicating that part of the energy used in accelerating the gas has also
heated it.

{\subsubsection{Mass, momentum and momentum rates}\label{sec-parameters}}
The mass in the outflows can be computed from the derived column densities
as,
\begin{equation}
M = [{\rm H}_2/^{12}{\rm CO}]~\mu_m~ \sum N_{\rm CO}~dA~~ ,\label{eq-mass}
\end{equation}
where $N_{\rm CO}$ is given by Eq. \eqref{eq-columnCO}, $\mu_m$ is the mean
molecular mass per H$_2$ molecule, [H$_2$/$^{12}$CO] the molecular hydrogen
to carbon monoxide abundance ratio, $dA$ the size of the emitting area in
an individual position, and the sum is over all the observed positions.

The derived masses of the SE-NW and NS flows are given in Columns 2-4 of
Table \ref{tbl-flows}. Column 2 gives the mass of the outflows in the
high-velocity (HV) ranges, computed using the column densities given in
Table \ref{param-vels}, $\mu_m=2.3\,m_H$, and $[{\rm H}_2/^{12}{\rm
    CO}]=10^4$.  To estimate the contribution to the mass from the
outflowing gas emitting in the same velocity range as the ambient cloud
gas, referred as the low-velocity (LV) outflow gas, we followed the
prescription of \citet{Margulis1985ApJ}. Using expression (A16) of
\citet{Bourke1997ApJ}, adopting as velocity boundary between the wing and
ambient emissions values of $-22$ \kms\ in the blue side and $-3$
\kms\ in the red side, we estimate that the mass of the low velocity flow
is 1.26 and 0.92\Msun\ for the SE-NW and NS flow, respectively.  The total
masses for the SE-NW and NS outflows are then 1.92 and 1.32 \Msun,
respectively.

The momentum of the gas can be estimated from the first moment of the line
emission. This procedure has the advantage of using the detailed
information of the spectra. The derivation of flow kinematic and dynamical
parameters by using moments of the spectrum has been described in detail by
\citet{Calvet1983ApJ}.  For this purpose we used the \co{3}{2} spectra
since it has the best signal-to-noise ratio, but note that using any of the
three transitions render consistent values within a 10\%.  In the low
velocity range we selected $-12.5$\ \kms, the ambient cloud velocity, as
the limiting value between what we consider blue or red-shifted gas.  The
derived momentums are given in Columns 5-7 of Table \ref{tbl-flows}.  They
were computed using the observed radial velocities, and thus correspond to
strict lower limits. To correct for inclination they should be multiplied
by a factor $(\cos i)^{-1}$ where $i$ is the angle between the flow and the
line of sight.

The average momentum rates can be estimated from the second moment of the
\co{3}{2} spectra, from the expression 
\begin{equation}
\label{eq-force}
 \dot{P}= [{\rm H}_2/^{12}{\rm CO}]~\mu_m~\sum_\textrm{ lobe} F( J,T_{\rm ex})\frac{\int \tau_v v^2 dv}{r_{\rm char}}~dA~, 
\end{equation}
where $F( J,T_{\rm ex})$ is the term that multiplies the integral in the
right hand side of Eq. \eqref{eq-columnCO}, and we select $J=2$ as
indicated previously.  The opacity $\tau_v$ can be estimated from
Eq. \eqref{eq-tauvsJ} in the optically thin limit.  The characteristic
radius of the flow, $r_{\rm char}$, is a sensitive parameter that regulates
the size of the flow and the time scale (after dividing it by the flow
velocity) over which the momentum has been deposited in the pre-stellar
core.  In Eq. \eqref{eq-force}, we consider $r_{\rm char}$ as the projected
radius, taken as the length subtended by 30\arcsec~(0.23 pc at 1.6 kpc) for
the high-velocity flow; and as the length subtended by 20\arcsec\ for the
low-velocity gas.  The momentum rates derived for the outflows, with no
correction for inclination, are given in cols. 8-10 of Table
\ref{tbl-flows}. A correction for inclination will require to multiply by
the factor $(\cos i)^{-2}\sin i$.

Toward each of the lobes of the SE-NW flow we observe wing emission 
at both blueshifted and redshifted velocities. This allows us to 
make an estimate of its inclination using the expression 
 \citep{Cabrit1990ApJ,Cabrit1986ApJ}
\[ \tan i = \frac{R+1}{R-1} \frac{1}{\tan \theta} ~~,\]
where $R$ is the ratio between maximum observed blueshifted and redshifted
velocities toward a single lobe, and $\theta$ is half of the total aperture
angle of the outflow.  We estimate the later from the 2-MASS $K_s$-band
image, obtaining a value of $\sim 23$\degs. Using this and the observed
values of R, we derive an inclination angle for the SE-NW outflow of
80\degs. Correction factors for the momentum and momentum rate are then 5.8
and 33, respectively.  The inclination-corrected momentum of the SE-NW flow
is $89 \Msun\ \kms$ and the inclination-corrected momentum rate (calculated
with Eq. \ref{eq-force}) is $3.0 \times 10^{-2}\Msun\,\kms$ yr$^{-1}$.

One of the principal problems of the previous estimation is that one
assumes that the momentum has been injected into the molecular gas
continuously, despite the molecular high velocity gas and the radio images
evidence that the emission comes from discrete episodes of mass ejection.
Therefore, the momentum rates calculated are time-averaged over the history
of mass ejections from the protostar, that probably includes long periods
of inactivity.

The momentum flux of the ionized jet reported by \citet{Guzman2010ApJ}
---corrected for 80\degs\ inclination--- is $\sim4.2\times10^{-4}$
\Msun~\kms\ yr$^{-1}$, while that of the associated SE-NW molecular flow
($3.0\times10^{-2}\Msun~\kms$\ yr$^{-1}$) is approximately 70 times
greater. Also, the observed momentum of the flow could not have been
deposited in the ambient gas by such a jet in less than $\sim$10$^5$ yr, a
time larger than the estimated life of the YSO.  A possible solution to
this inadequacy of the jet as the driving source of the molecular outflow
could be that the protostellar jet is only partially ionized, being its
mass and momentum loads greater than the estimated ones.

{\subsection{Further tracers of shocks}\label{sec-shocks}}

\begin{figure}
\centering
\includegraphics[angle=-90,width=.6 \textwidth ]{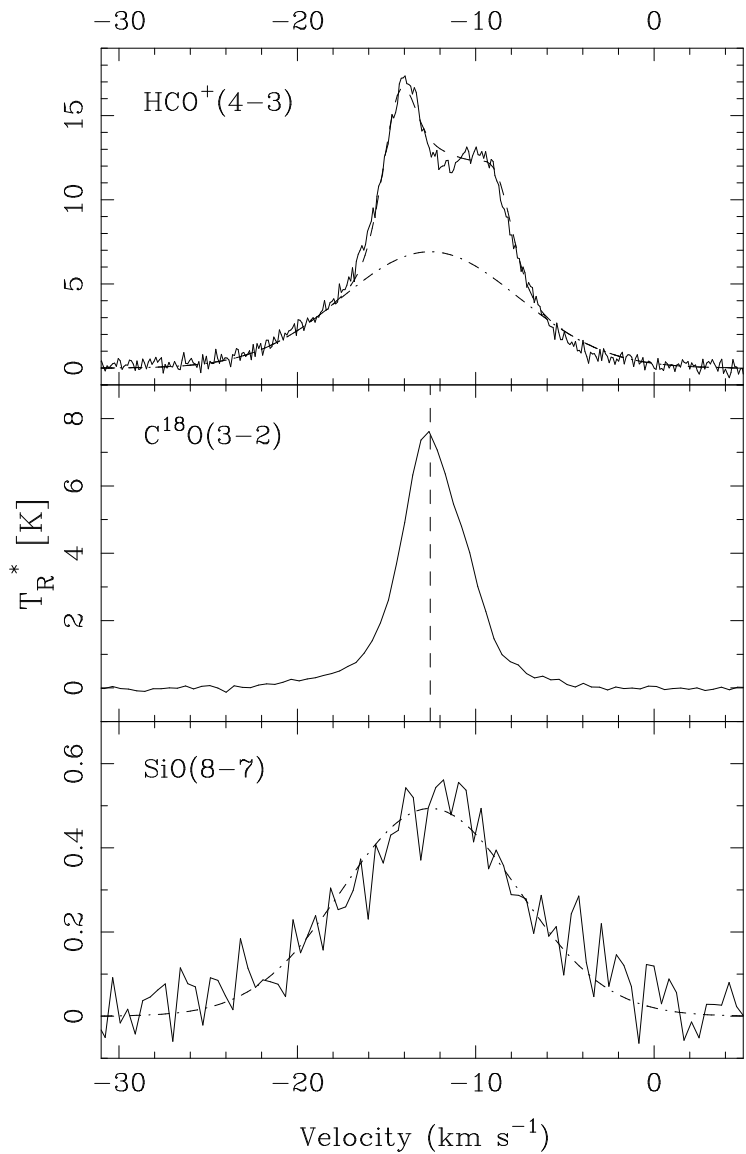}
\caption{\baselineskip0.0pt Spectra observed toward
the peak position of IRAS 16562$-$3959. \textit{Top:} \hcop{4}{3} spectrum.
The dashed line shows the best-fit using a collapsing envelope model 
from \citet{Myers1996ApJ} plus an outflow component (dot-dashed line). 
  \textit{Middle:} \cdo{3}{2} spectrum. The dashed-line indicates the 
  adopted ambient cloud velocity of $-12.5~ \kms$.
 \textit{Bottom:} \sio{8}{7} spectrum. The
  dot-dashed line is not a fit but shows the outflow profile fitted to the
  \hcop{4}{3} spectrum scaled by $1/14$. \label{fig-hcomas}}
\end{figure}

\begin{figure}
\includegraphics[angle=0,height=.8\textheight ]{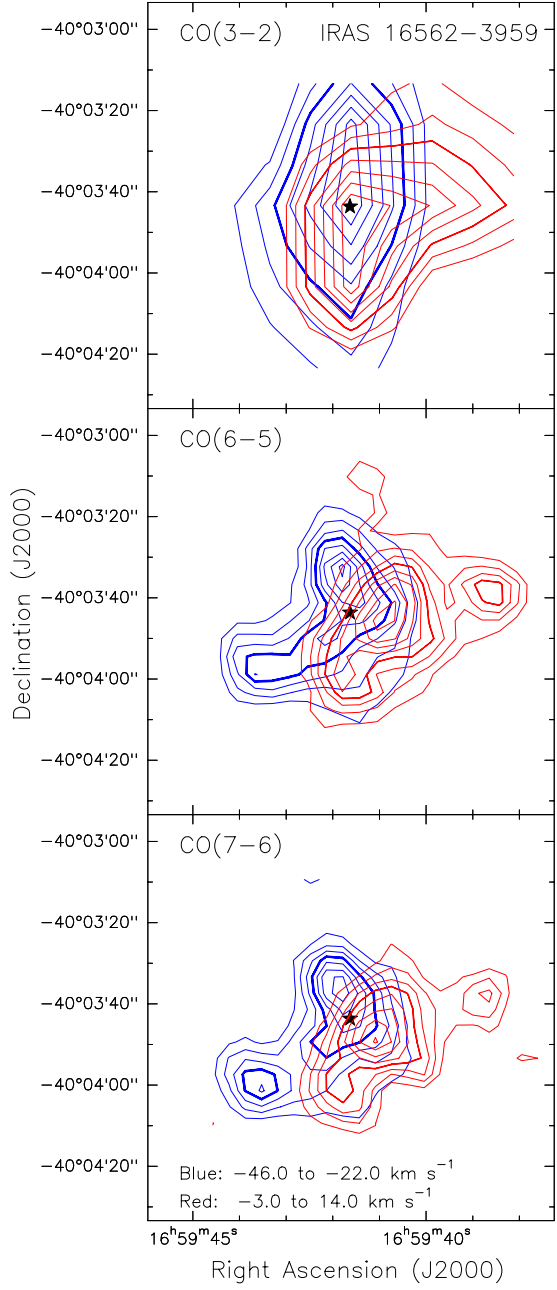}
\caption
{\baselineskip0.0pt Contour maps of the velocity integrated CO line wing
  emission towards IRAS 16562$-$3959. Blue lines represent emission integrated
  over the velocity range $-46< v_{lsr}< -22$ \kms, which is
  blueshifted with respect to the ambient velocity of $-12.5$ \kms, and red
  dashed lines emission integrated over the velocity range
  $-3<v_{lsr}<14$ \kms, which is redshifted with respect to the
  ambient velocity. The star marks the position of the jet source. Contour
  levels are 20, 30, 40, 50, 60, 70, 80, and 90\% of the peak emission.
  Top: \co{3}{2} emission. Peak blueshifted emission: 49.0 K
  \kms.  Peak redshifted emission: 26.4 K \kms.  Middle:
  \co{6}{5} emission. Peak blueshifted emission: 85.2 K \kms.
  Peak redshifted emission: 49.6 K \kms.  Bottom: \co{7}{6}
  emission. Peak blueshifted emission: 85.1 K \kms.  Peak redshifted
  emission: 54.3 K \kms.
\label{fig-co-outflows}}
\end{figure}

\begin{figure}
\includegraphics[angle=0,width=1 \textwidth ]{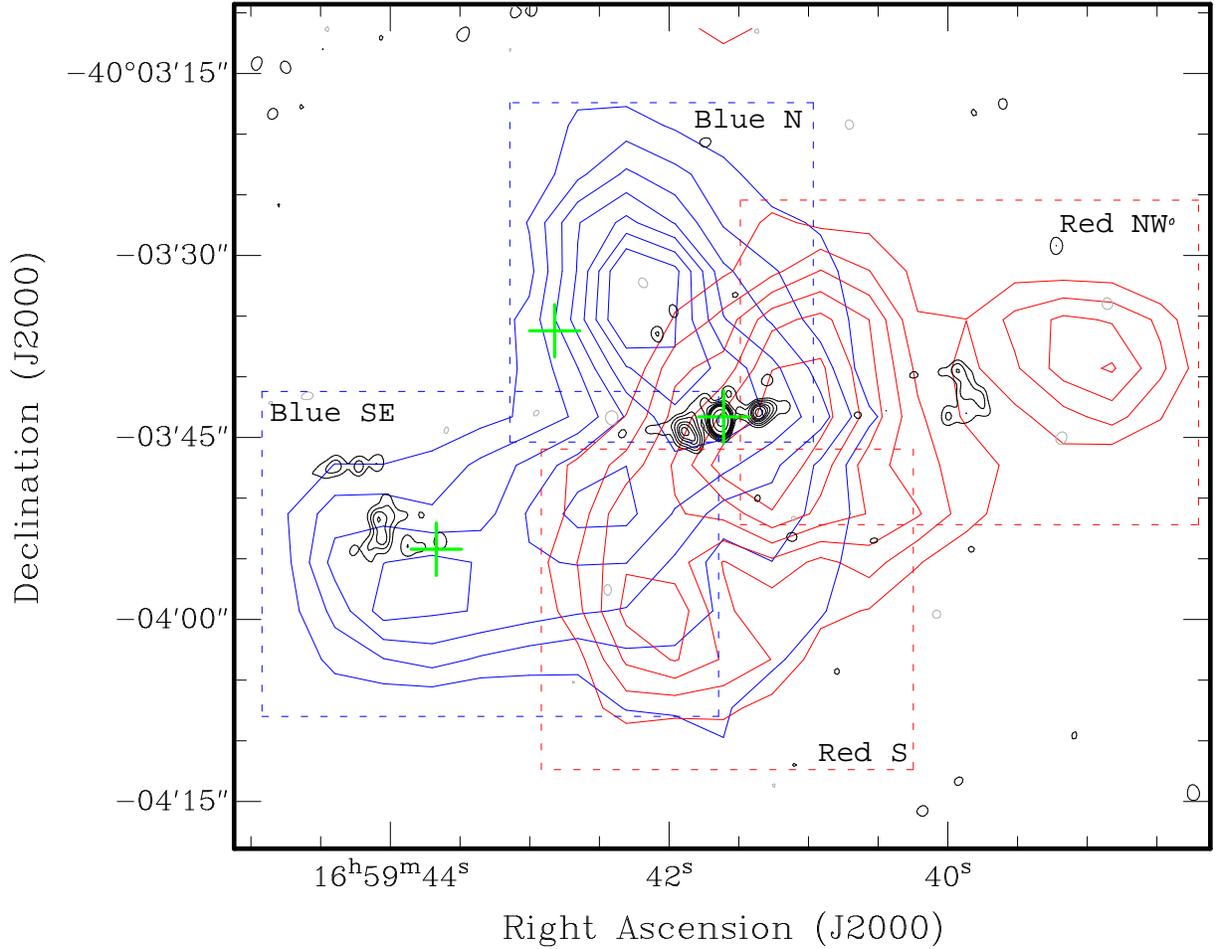}
\caption{\baselineskip0.0pt Map of the \co{6}{5} velocity integrated
  blueshifted (blue contours) and redshifted (red contours) wing emission,
  as in Fig. \ref{fig-co-outflows}, overlayed with a map of the 8.6 GHz
  emission (black contours) showing the string of radio emission described
  in \citet{Guzman2010ApJ}.  The red and blue dashed-line boxes mark the
  regions of the sky where the CO emission was integrated to determine the
  parameters of the red and blue lobes, respectively. The green crosses marks
  the position of the OH masers.
\label{fig-outjet}}
\end{figure}

\begin{figure}
\hspace*{-2 cm}\includegraphics[angle=-90,width=1.2 \textwidth ]{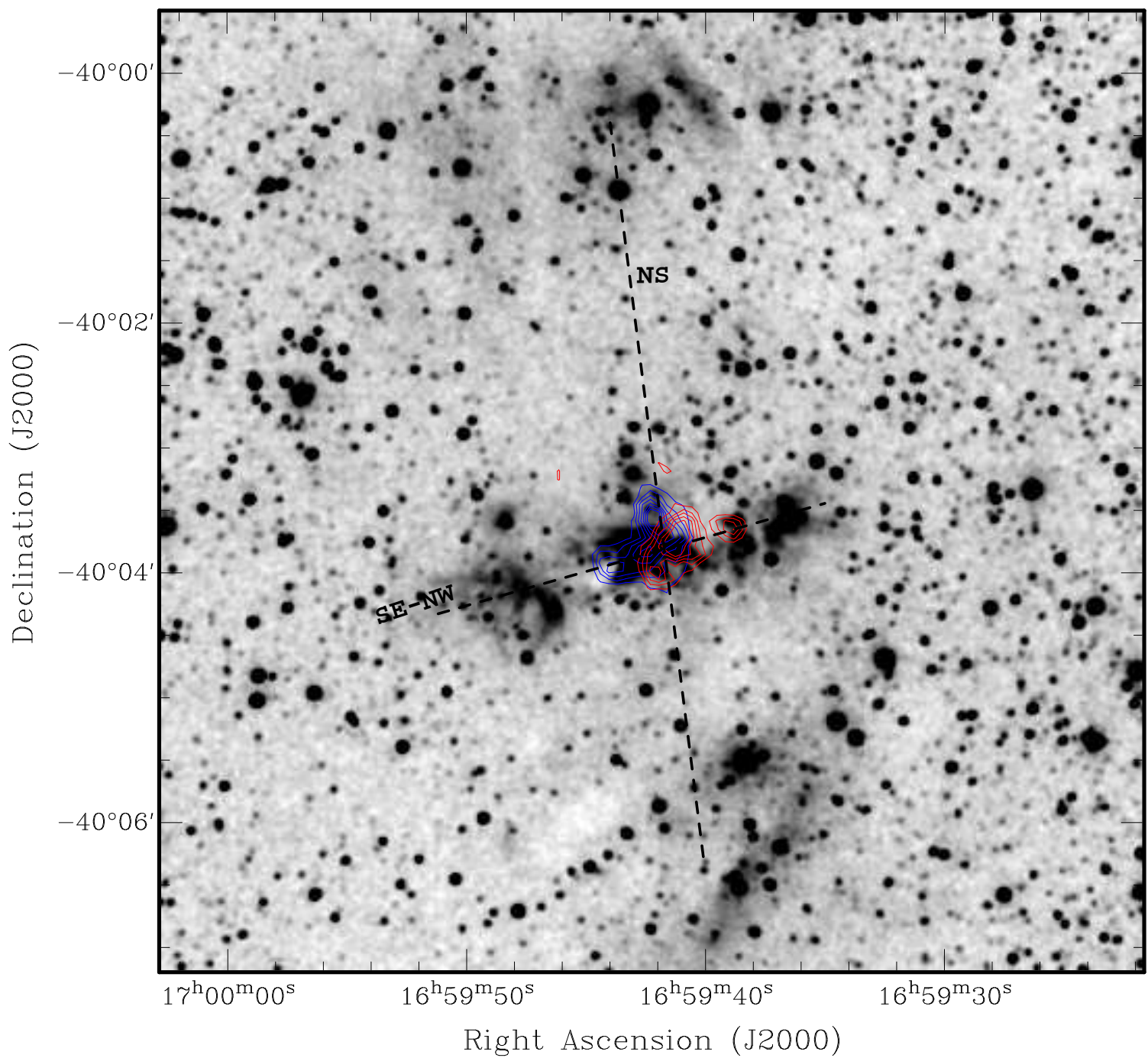}
\caption{\baselineskip0.0pt
\emph{Greyscale}: $K_s$-band 2MASS emission. Overlaid are 
contours of the blueshifted (blue contours) and redshifted (red contours)  
\co{6}{5} emission. The approximate directions of the SE-NW and NS flows are 
indicated by dashed lines.
\label{fig-2mass}}
\end{figure}

\begin{figure}\centering
\includegraphics[angle=-90,width=.8 \textwidth ]{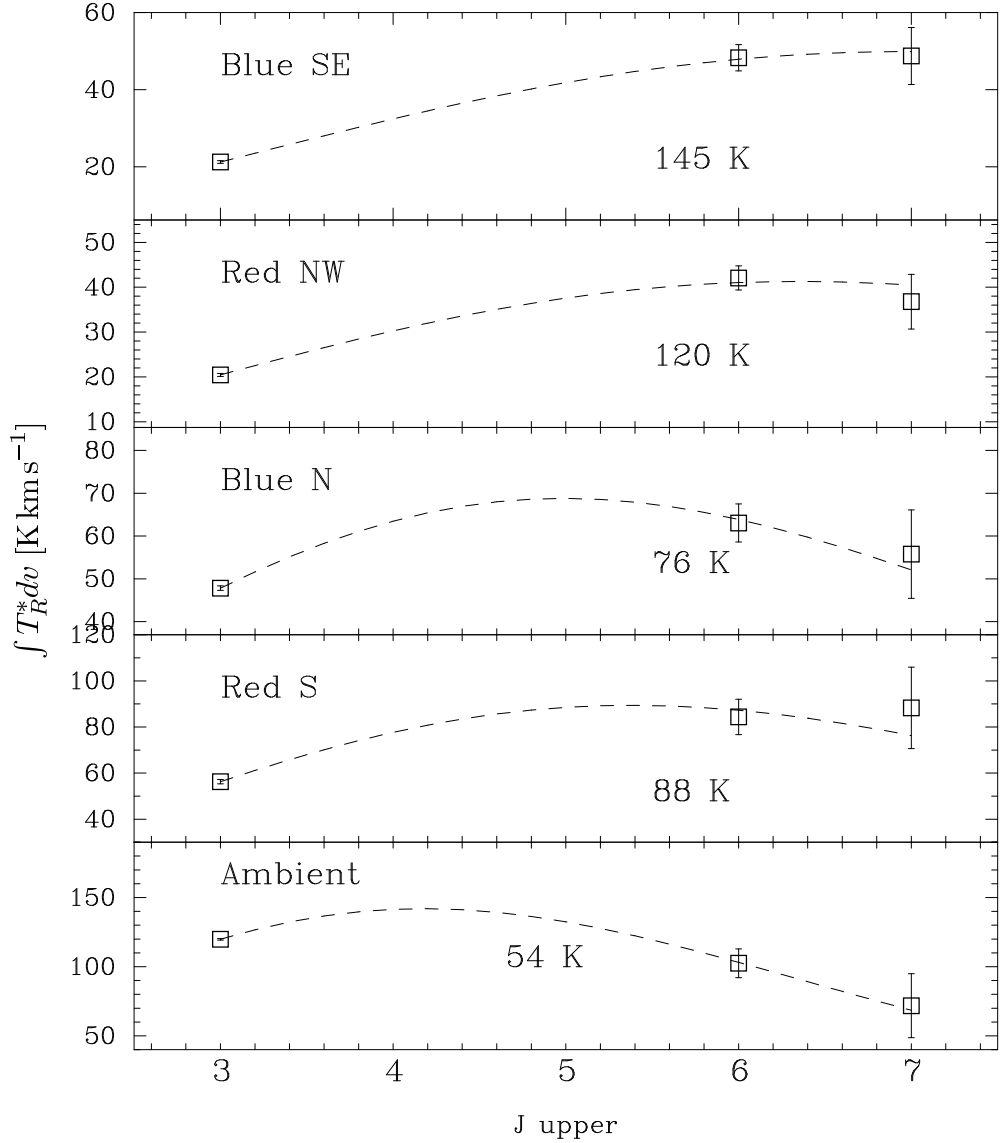}
\caption{\baselineskip0.0pt Velocity integrated CO emission versus upper
  rotational quantum level $J$. The range of velocity integration for the
  different features is indicated in Column 2 of Table \ref{param-vels}.
  The outflow emission from the lobes has been averaged over the zones
  indicated in Fig. \ref{fig-outjet}. Dashed lines represent fits to the
  data using Eq. \eqref{eq-columnCO} and optically thin conditions.  The
  derived column densities and excitation temperatures are shown in Table
  \ref{param-vels}, and the latter are also indicated in each panel.
  \textit{Top:} Wing emission from the blueshifted lobe of the SE-NW
  outflow.  \textit{Middle-top:} Wing emission from the redshifted lobe of
  the SE-NW outflow.  \textit{Middle:} Wing emission from the blueshifted
  lobe of the NS outflow.  \textit{Middle-bottom:} Wing emission from the
  redshifted lobe of the NS outflow.  \textit{Bottom:} Ambient cloud
  emission, measured from spectrum observed at offsets
  $\Delta\text{RA}=+40\arcsec,\Delta\text{Dec}=-40\arcsec$.
\label{fig-vels}}
\end{figure}

In addition to the high-velocity CO gas and its high excitation
temperature, further evidence for the presence of shocks is provided by the
detection of SiO emission.  SiO is thought to be generated when strong
shocks pass through dense molecular gas disrupting dust grains
\citep{Hartquist1980ApJ,Caselli1997AA}. Therefore, we expect that SiO
emission should be associated with the outflowing gas. This suggestion is
strongly supported by the almost identical shape of the profile of the high
velocity outflowing gas emission observed in the \hcop{4}{3} line and the
\sio{8}{7} profile (see next section and bottom panel of
Fig. \ref{fig-hcomas}).

The observed velocity integrated SiO emission is $\sim4.6$ K \kms.
Assuming the SiO line is emitted in optically thin conditions with an
excitation temperature of 75 K, we derive a column density of SiO of
7.8$\times 10^{12}$ cm$^{-2}$.  Assuming $T_{\rm ex}=100$ K only changes
this value by 3\%. The critical density for the \sio{8}{7} transition is
$\sim 10^8$ cm$^{-3}$, considerable larger than the ambient density,
indicating that thermal collisions with ambient H$_2$ is not the main
mechanism for excitation to this rotational state.  The good correlation
between the SiO emission and the high velocity component suggests that this
mechanism is shock excitation of the gas.

There are three OH masers reported in the literature toward the region
\citep{Caswell1998MNRAS297215,Caswell2004MNRAS}. Figure \ref{fig-outjet}
shows their positions with green crosses. The closest to the peak of the SE
blue-shifted lobe corresponds to a 1720 MHz OH maser, which has  an LSR 
velocity of $-24.5$\ \kms. The 1720 MHz OH masers are usually associated with
shocked gas \citep{Brogan2007IAUS}, and the location and velocity of this
particular one is consistent with it being pumped by the shock produced by
the high velocity blueshifted flow. \citet{Caswell2004MNRAS} reported
Zeeman splitting on this maser implying a $\sim10$ mG magnetic
field, the latter probably enhanced through shock compression.  We also
note that the excitation temperature of the outflowing gas ---derived in \S
\ref{sec-NandTex}--- is above the minimum temperature required to pump this
maser collisionally \citep[${\rm T}\approx 90$
  K,][]{Elitzur1976ApJ,Hoffman2003ApJ}.  We conclude that this maser is
associated with the IRAS 16562$-$3959 core and to the shocked gas in the SE-NW
outflow, and not located farther away at $\sim 2.5$ kpc as previously
suggested by \citet{Caswell2004MNRAS}.

The other two masers correspond to 1665/1667 MHz OH masers: one is
coincident with the central jet source ($V_\text{OH}=-12.7\,$\kms) and the
other, located about 15.6\arcsec\ NE from the jet
($V_\text{OH}=-15.5$\ \kms), is associated with a mid infrared source in
the field, unseen in 2MASS but conspicuous in  \emph{Spitzer} 8.0\um 
band (see Fig. 4 from \citealt{Guzman2010ApJ}).
 
{\subsection{Infall of molecular core}\label{sec-collap}}

To derive infall parameters we modeled the \hcop{4}{3} profile using the
simple analytic model of contracting clouds of \citet{Myers1996ApJ}.  We first
subtracted a broad Gaussian profile, shown as a dash-dotted line in the
upper panel of Fig. \ref{fig-hcomas}, to take into account the contribution
of the high velocity outflow. We note that the shape of the subtracted
Gaussian is very similar to the profile of the SiO($8\rightarrow7$)
emission which is expected to come mainly from the outflowing gas.  This is
illustrated in the bottom panel of Fig.  \ref{fig-hcomas}, which shows the
SiO($8\rightarrow7$) spectrum and superimposed the broad Gaussian profile
fitted to the \hcop{4}{3} emission (scaled by a factor of 1/14).

The simple model can very well fit the observed profile (see top panel of
Fig.  \ref{fig-hcomas}).  The parameters of the best fitting collapse model
are: an infall velocity ($V_{\rm in}$) of 0.35 $\kms$, a velocity dispersion
($\sigma$) of 1.54 $\kms$, an optical depth ($\tau_0$) of $3.8$, a kinetic
temperature of $32$ K, and a rest-frame velocity of the collapsing envelope
of $-11.4$ \kms.  This velocity is about 1 \kms\ greater than the ambient
cloud velocity adopted here, but we note that \citet{Bronfman1996AAS} also
reported a velocity of $-11.6$ \kms\, in the CS(2$\rightarrow$1) transition
detected towards this source.

In order to estimate the envelope accretion rate we consider the mass
distribution model for the parent cloud derived from dust emission in
\citet{Guzman2010ApJ}.  Assuming that the extent of the contracting gas
along the line of sight is comparable to the size of the cloud of about 0.7
pc FWHM (1.5 \arcmin\ at 1.6 Kpc) and using the derived values of the
infall speed, molecular density, and core size, we obtain a mass infall
rate $\dot{M}_{in}$ of $\sim3.7\times10^{-4}$\Msun\ yr$^{-1}$, a value one
order of magnitude below the estimation from the SED fitting reported by  
\citet{Guzman2010ApJ}.
The value derived for the infall velocity is also about one order of 
magnitude below the free-fall velocity expected for the entire cloud, 
suggesting a retarded collapse.

\section{SUMMARY}

We undertook molecular line observations toward the IRAS 16562$-$3959 massive
and dense core in three CO transitions: CO(3$\rightarrow$2),
CO(6$\rightarrow$5), and CO(7$\rightarrow$6), using APEX, with angular
resolutions ranging from $\sim$8\arcsec\ to $\sim$17\arcsec. Additionally,
\hcop{4}{3}, \cdo{3}{2} and \sio{8}{7} observations were performed towards
the center of the core. The main results and conclusions presented in this
paper are summarized as follows:
\begin{enumerate}

\item{High velocity molecular gas was detected in the three CO transitions,
 spanning a range in radial velocity of $\sim 60~\kms$.}
\item{The morphology of the high velocity emission, best shown in the
  \co{6}{5} and \co{7}{6} lines, is quadrupolar. We conclude that this
  morphology is due to the presence of two collimated bipolar outflows, one
  lying in the SE-NW direction and another in the N-S direction.  The more
  extended bipolar flow is in the SE-NW direction, consistent with being
  excited by the ionized jet detected toward this MYSO
  \citep{Guzman2010ApJ}.  Extended $K_s$-band emission probably tracing
  excited H$_2$-2.12\,\um\ is also associated with the SE-NW flow.}
\item{From an analysis of the emission in the three CO lines we derive that
  the excitation temperature of the high-velocity gas, assumed optically
  thin, is high, with values of $\sim140$ in the lobes of the SE-NW outflow
  and $\sim90$ in the lobes of the N-S outflow.  This suggests that the
  excitation of the gas is related with the acceleration mechanism, namely,
  shock-induced acceleration.}
\item{The total mass in the SE-NW and NS outflows are, respectively, 1.9
  and 1.3\Msun. The momentum and momentum rates derived for the SE-NW
  outflow, corrected for an inclination angle of 80\degs, are
  89\Msun\ \kms\ yr$^{-1}$ and $3.0\times 10^{-2}$ \Msun\ \kms\ yr$^{-1}$,
  respectively. These values are characteristics of flows driven by young
  massive stellar objects with high luminosities.}
\item{The molecular core in which the outflow is embedded presents evidence
  of being in gravitational contraction as shown by the blue asymmetric
  peak seen in the \hcop{4}{3} transition observations. The derived mass 
  infall rate is of the order of a few times $10^{-4}$\Msun~yr$^{-1}$ and 
  the infall velocity is $\sim0.35$~\kms,
  which is a factor of $\sim10$ below the velocity expected from a free-fall
  collapse. This implies that the collapse has been retarded
  and not gravitationally dominated.  }
\end{enumerate}

\acknowledgments

The authors gratefully acknowledge support from CONICYT through projects
FONDAP No. 15010003 and BASAL PFB-06.  This publication is partly based on
data acquired with the Atacama Pathfinder Experiment (APEX). APEX is a
collaboration between the Max-Planck-Institut f\"ur Radioastronomie, the
European Southern Observatory, and the Onsala Space Observatory.  This
publication makes use of data products from the Two Micron All Sky Survey,
of the GLIMPSE-Spitzer database, and of the
Red \emph{MSX} Source survey database at www.ast.leeds.ac.uk/RMS.

\newpage

\begin{deluxetable}{ccccccccc}
  \tablewidth{0pt}
  \tablecaption{Observational Parameters \label{tbl-obs}}
  \tablehead{
    \colhead{Line} & \colhead{Frequency} & Tel.&\colhead{Beam} &\colhead{$\eta_{mb}$}&\colhead{Map}&\colhead{Spacing}&
\colhead{$\Delta v$}&\colhead{Noise} \\
   \colhead{} & \colhead{[GHz]} & \colhead{}&\colhead{[\arcsec]} &\colhead{}&\colhead{}&
    \colhead{[\arcsec]}
&\colhead{[\kms]}&\colhead{[K]}
  }
  \startdata
\co{3}{2}  &  345.796 & APEX &$17.3$ & $0.73$ & $5\times5$   & 20\arcsec& $0.423$ & $0.1\phn$\\
\co{6}{5}  &  691.627 & APEX &$ 8.8 $& $0.47$ & $29\times21$ & \phn4\arcsec  & $0.318$ & $0.5\phn$\\
\co{7}{6}  &  806.898 & APEX &$ 7.7 $& $0.43$ & $27\times19$ & \phn4\arcsec  & $0.272$ & $1.0\phn$\\
\sio{8}{7} &  347.331 & APEX &$17.$  & $0.73$ & $3\times3$   & 20\arcsec& $0.423$ & $0.05$\\
\hcop{4}{3}& 356.734  & ASTE &$22$ & $0.7\phn$& Central Pos. &   ---    &  0.11\phn & 0.1\phn\\
\cdo{3}{2} & 329.331  & ASTE &$23$ & $0.7\phn$& Central Pos. &   ---    & 0.43\phn & 0.05 \\
\enddata
\end{deluxetable}

\begin{deluxetable}{rlcr}
  \tablewidth{0pt}
  \tablecaption{ Parameters of High-Velocity and Ambient Gas\label{param-vels}}
  \tablehead{\colhead{Lobe}&\colhead{V range}&\colhead{N$_\text{CO}$}&\colhead{T$_\text{ex}$} \\ 
\colhead{}&\colhead{(\kms)}&\colhead{(10$^{16}$ cm$^{-2}$)}&\colhead{(K)} }
  \startdata
Blue SE  &$[-46.0,-22]$ &\phn8.72  &   145 \\
Red NW   &$[-3,14.0]$   &\phn6.06 &   120 \\
Blue  N  &$[-46.0,-22]$ &\phn5.70 &\phn76 \\
Red  S   &$[-3,14.0]$   &\phn3.37 & \phn88 \\
Ambient  &$[-22,-3]$    &11.30 &\phn54 \\
\enddata
\end{deluxetable}

\begin{planotable}{rccccccccc}
  \tablewidth{0pt}
  \tablecaption{Outflow Parameters \tablenotemark{a}\label{tbl-flows}}
  \tablehead{ & \multicolumn{3}{c}{Mass} & \multicolumn{3}{c}{Momentum\tablenotemark{b}} 
  & \multicolumn{3}{c}{Momentum rate\tablenotemark{b}} \\
  \multicolumn{1}{c}{Outflow} & \multicolumn{3}{c}{\small (\Msun)} 
  & \multicolumn{3}{c}{\small (\Msun\,\kms)} & 
  \multicolumn{3}{c}{\small ($10^{-4}\Msun\,\kms\,\text{yr}^{-1}$)}\\
 & HV  & LV & Total &HV & LV & Total &HV & LV & Total }
  \startdata
\cutinhead{SE-NW outflow}
Blue SE & 0.39 & 0.61 & 1.00 & 5.8 & 2.9 & \phn9.7 & 4.3 & 1.2 & 5.5 \\
Red NW  & 0.27 & 0.65 & 0.92 & 3.5 & 3.1 & \phn6.6 & 2.2 & 1.3 & 3.5 \\
Total   & 0.66 & 1.26 & 1.92 & 9.3 & 6.0 &    15.3 & 6.5 & 2.5 & 9.0 \\
\cutinhead{NS outflow}
Blue N  & 0.25 & 0.49 & 0.74 & 3.4 & 2.4 & 5.8 & 3.3 & 1.0 & 4.3 \\
Red S   & 0.15 & 0.43 & 0.58 & 1.9 & 2.0 & 3.9 & 1.7 & 0.8 & 2.5 \\
Total   & 0.40 & 0.92 & 1.32 & 5.3 & 4.4 & 9.7 & 5.0 & 1.8 & 6.8 \\ 
\enddata
\tablenotetext{a}{HV and LV denotes parameters of the high-velocity 
and low-velocity flows, derived as described in \S \ref{sec-parameters}.}
\tablenotetext{b}{Not corrected for inclination.}
\end{planotable}


\clearpage 

\bibliographystyle{apj}
\bibliography{bibliografia}

\newpage

\end{document}